%
\input phyzzx.tex
\tolerance=1000
\voffset=-0.3cm
\hoffset=1.0cm
\sequentialequations
\def\rl{\rightline}

\def\t1{{\tilde 1}}

\def\NPB#1#2#3{Nucl. Phys. B {\bf#1} (19#2) #3}
\def\PLB#1#2#3{Phys. Lett. B {\bf#1} (19#2) #3}

\def\PRT#1#2#3{Phys. Rep. {\bf#1} (19#2) #3}

\def\IJMP#1#2#3{Int. J. Mod. Phys. A {\bf#1} (19#2) #3}

\REF\SSB{V. Kaplunovsky and J. Louis, \PLB{306}{93}{269}; A. Brignole, 
L. Ibanez and C. Munoz, \NPB{422}{94}{125}.}
\REF\GCM{H. P. Nilles, \PLB{115}{82}{193}; 
J. P. Deredinger, L. E. Ibanez and H. P. Nilles, \PLB{155}{85}{65};
M. Dine, R. Rohm, N. Seiberg and E. Witten, \PLB{156}{85}{55}.}
\REF\KRA{N. V. Krasnikov, \PLB{193}{87}{37}; J. A. Casas, Z. Lalak, C. Munoz and G. G. Ross, \NPB{347}{90}{243}.}
\REF\CAR{B. de Carlos, J. Casas and C. Munoz, \NPB{399}{93}{623}.}
\REF\IBA{A. Font, L. Ibanez, D. Lust and F. Quevedo, \PLB{245}{90}{401}.}
\REF\BG{P. Binetruy and M. K. Gaillard, \NPB{358}{91}{121}.}
\REF\TSD{A. Giveon, M. Porrati and E. Rabinovici, Phys. Rep. {\bf 244} (1994)
77 and references therein.}
\REF\FER{S. Ferrara, L. Girardello and H. P. Nilles, \PLB{125}{83}{457};
S. Ferrara, N. Magnoli, T. Taylor and G. Veneziano, \PLB{245}{90}
{409}.}
\REF\LT{D. Lust and T. Taylor, \PLB{253}{91}{335}.}
\REF\SQCD{D. Amati, K. Konishi, Y. Meurice, G. C. Rossi and G. Veneziano,
\PRT{162}{88}{169}.}
\REF\THR{V. Kaplunovsky, \NPB{307}{88}{145}; L. Dixon, V. Kaplunovsky and 
J. Louis, \NPB{355}{91}{649}.}
\REF\DSW{M. Dine, N. Seiberg and E. Witten, \NPB{289}{87}{589};
J. J. Atick, L. J. Dixon and A. Sen, \NPB{292}{87}{109}
 S. Cecotti, S. Ferrara and M. Villasante, \IJMP{2}{87}{1839}.}
\REF\EDI{E. Halyo, \PLB{343}{95}{161}.}
\REF\EA{A. E. Faraggi and E. Halyo, preprint IASSNS-HEP-94/17, hep-ph/9405223.}
\REF\CKM{A. E. Faraggi and E. Halyo, \PLB{307}{93}{305}; \NPB{416}{94}{63}; 
E. Halyo, \PLB{332}{94}{66}.}
\REF\RP{E. Halyo, Mod. Phys. Lett. {\bf A9} (1994) 1415.}

\singlespace
\rl{SU-ITP-96-4}
\rl{\today}
\pagenumber=0
\normalspace
\medskip
\titlestyle{\bf{Dilaton Supersymmetry Breaking}}
\smallskip
\author{Valerie Halyo{\footnote*{address: 2680 Fayette Dr. Mountain View CA}} and 
Edi Halyo$^1${\footnote\dagger{e--mail address: halyo@leland.stanford.edu}}}
\smallskip
\centerline {$^1$Department of Physics} 
\centerline {Stanford University}
\centerline {Stanford, CA 94305}
\vskip 2 cm
\titlestyle{\bf ABSTRACT}

We argue that dilaton supersymmetry breaking in string derived
supergravity requires an effective superpotential which is not 
separable as a function of the dilaton times a function of the 
moduli. We show that in a simple model with hidden sector matter 
condensation and a dilaton independent term one can easily 
obtain $|F_S| \not=0$. For a wide range of realistic model 
parameters $|F_S|>>|F_T|$ and supersymmetry is mainly broken 
in the dilaton direction.

\singlespace
\vskip 0.5cm
\endpage
\normalspace

\centerline{\bf 1. Introduction}

It is well-known that supersymmetry (SUSY) breaking in the dilaton 
direction (i.e. due to
a nonzero dilaton F--term) has a number
of attractive features. In this scenario, due to the universal dilaton
coupling to all matter, the soft SUSY breaking scalar and gaugino masses as
well as the A--terms are universal[\SSB]. In addition, these only depend on one
parameter, the gravitino mass, $m_{3/2}$, which parametrizes the 
nonperturbative SUSY breaking effects. The results for the soft SUSY breaking
parameters are the universal boundary conditions often used in
supergravity 
(SUGRA) models. In this case, the parameter space is reduced by a large 
amount which makes
the low--energy calculations much more predictive. Moreover, the universal
soft scalar masses obtained in this scenario eliminate the danger of flavor
changing neutral currents at the weak scale even after running effects are
taken into account [\SSB]. Thus, it is highly desirable to obtain dilaton SUSY breaking
in string derived SUGRA models.

In conventional SUSY breaking mechanisms obtained by hidden sector gaugino
condensation[\GCM] in string derived SUGRA models there are two problems related to the
dilaton. First, the resulting nonperturbative effective superpotentials give 
effective scalar potentials which are not stable in the dilaton direction; i.e.
the dilaton runs to infinity and there is no stable vacuum. (A possible
solution to this problem is given by racetrack models[\KRA,\CAR] which have more than
one condensing hidden gauge group but these require fine tuning of the
coefficients.) In this letter, we do not try to solve this problem. 
Instead, we add
a dilaton independent term to the nonperturbative superpotential which
stabilizes the dilaton potential[\IBA]. The origin of such a
term may be the condensation of the antisymmetric field tensor present in
all string models[\BG]. In any case, this term parametrizes the effects
of the unknown nonperturbative mechanism which stabilizes the dilaton 
potential.

Second, even though nonzero VEVs for moduli F--terms ($F_T$) are very easy
to obtain in this
scenario,  the dilaton F--term $F_S$ always vanishes. This has been shown 
for hidden sectors with a pure gauge group 
with or without a dilaton independent term[\IBA]. 
The reason for the vanishing  
$F_S$ in the models considered so far can be traced back to the
separability  of the
nonperturbative superpotential. That is $W_{np}=g(S)h(T)$ for the cases
considered so far; the superpotential is a product of a function of 
the dilaton $S$ and a function of the overall modulus $T$. (We consider
only the overall modulus case but the generalization to the 
multimodulus case is trivial.)
In this case, it can be shown that the minimum of the effective potential
in the dilaton direction occurs where $F_S=0$, i.e. there is no dilaton SUSY
breaking[\CAR,\IBA]. Therefore, in order to have dilaton breaking one should look at 
effective superpotentials which are not separable.

In this letter, we investigate the case with a nonseparable superpotential.
This is obtained by considering a model with hidden matter in 
the vector representation of the hidden gauge group and a dilaton 
independent term in the effective superpotential which stabilizes the 
dilaton.  We assume that target space duality[\TSD]
is unbroken by nonperturbative effects such as condensation in the 
hidden sector.
Invariance under target space duality fixes 
the moduli dependence of different terms in the superpotential.
We show that generically $F_S \not=0$ in addition to $F_T \not=0$. The ratio
$|F_S|/|F_T|$ depends on the parameters of the model such as the hidden
gauge group, the hidden matter content, their masses and the 
coefficient of the dilaton independent term.
We find that for a large range of realistic values of parameters, 
$S \sim O(1)$ and
$|F_S|>>|F_T|$ which gives dominantly dilaton SUSY breaking. In addition by
properly choosing the magnitude of the constant term we obtain 
$|F_S| \sim 10^{-15} M_P^2$ which gives a $TeV$ scale soft scalar masses.
The cosmological constant is $O(m_{3/2}^2 M_P^2)$ and negative as usual.
We make no attempt to solve this problem but note that it can be made to
vanish by fine tuning the constant term which stabilizes the dilaton.

\bigskip
\centerline{\bf 2. Separable and nonseparable superpotentials}

In all gaugino condensation scenarios considered so far the effective
nonperturbative superpotential is a function of the dilaton times a function
of the moduli,
$$W_{np}=g(S)h(T) \eqno(1)$$
For example for a hidden pure $SU(N)$ gauge group $W_{np}$ is given by[\FER,\LT]
$$g(S)=-Nexp(-32 \pi^2S/N) \eqno(2) $$
and
$$h(T)=(32 \pi^2 e)^{-1}\eta(T)^{-6} \eqno(3)$$
Here $\eta(T)$ is the Dedekind eta function whose
power is determined by target space duality invariance.
In this case,
one can add a dilaton independent term $c\eta(T)^{-6}$ to $W_{np}$
(where $c$ is a constant) 
in order to stabilize the dilaton potential[\IBA].  
$W_{np}$ is still separable
because both $g(S)$ and $c$ have the same (i.e. vanishing)
modular weights.
The dilaton and moduli F--terms are obtained from ($i=S,T$)
$$F_i=e^{K/2}(W_i+K_iW) \eqno(4) $$
and are given by
$$F_S=e^{K/2}(g_Sh(T)+K_Sg(S)h(T)) \eqno(5)$$
and
$$F_T=e^{K/2}(g(S)h_T+K_Tg(S)h(T)) \eqno(6)$$
with the Kahler potential
$$K(S, \bar S,T, \bar T)=-log(S+ \bar S)-3log(T+ \bar T) \eqno(7)$$
The effective nonperturbative scalar potential is given by
$$V_{eff}=|F_i|^2 G_{i \bar i}^{-1}-3e^K|W|^2 \eqno(8)$$
where $G=K+log|W|^2$ . 
Minimizing the effective scalar potential obtained by using Eqs. 
(1-8) one finds that there is a minimum
at $S$ which satisfies $g(S)-2S_R g_S=0$.
From Eq. (5) we see that this is exactly the condition for a vanishing dilaton
F--term, $F_S=0$. Thus we find generically that if the superpotential
is separable ${\partial V/ \partial S} \prop F_S$.
(There is another condition for an extremum of $V_{eff}$ but this 
corresponds to a maximum[\CAR].)

One possible way to obtain a nonseparable superpotential is to include 
effects of hidden matter condensation in $W_{np}$. Hidden sectors of
string models generically contain matter in vector--like representations
of the hidden gauge group. When the hidden gauge group condenses at
a scale $\Lambda_H \sim M_P exp(8\pi^2/bg^2)$ hidden matter condensates 
($\Pi$) form in addition to gaugino condensates ($Y^3$). 
Here $b$ is the coefficient of
the $\beta$--function and $g$ is the coupling constant at $M_P$ which is
about $0.7$. Integrating out $\Pi$ and $Y^3$ one
obtains $W_{np}$ as a function of the dilaton and moduli and $det A$
where $A$ is the hidden matter mass matrix which has to be nonsingular
for a stable vacuum[\SQCD]. 
In order to stabilize the dilaton we can add a dilaton independent 
term $c\eta(T)^{-6}$ to $W_{np}$ as mentioned above.
One can choose $c$ so that the dilaton VEV is
$O(1)$ in order to get $g^2 \sim 1/2$ for succesfull gauge coupling 
unification[\THR].  This term can arise from the condensation of the antisymmetric
tensor which is present in all string models[\BG].  In any case, we can view it 
as a parametrization of the unknown 
nonperturbative physics which stabilizes the dilaton.
As we will see, this new  term renders $W_{np}$ nonseparable
if it appears in a model with hidden matter condensation.

The effective nonperturbative superpotential for an $SU(N)$ gauge group
with $M$ matter
multiplets in the vector representations $N+ \bar N$ is given by[\LT]
$$W_{np}=g(S)h(T)[det A]^{1/N}+c k(T) \eqno(9)$$
where
$$g(S)=-Nexp(-32 \pi^2 S/N) \eqno(10)$$
and
$$h(T)=(32 \pi^2 e)^{M/N-1}\eta(T)^{2M/N-6} \eqno(11)$$
with the determinant of the hidden mass matrix $detA=k \phi^s \eta(T)^t$ [\EDI].
$k$ is a constant of $O(1)$, $\phi$ denote generic gauge singlet fields 
with modular weights $-1$ [\RP]and whose VEVs 
give hidden matter mass[\EA]. 
The powers $s$ and $t$ are models dependent but positive and of
$O(5-10)$.  The powers of $\eta(T)$ in $h(T), k(T)$ and $detA$ are determined from
target space duality invariance.  $k(T)=\eta(T)^{-6}$ since $c$ has vanishing 
modular weight. The power of $\eta(T)$ in $h(T)$ arises from the $M$ multiplets of
hidden matter with modular weights $-1$. $detA$ has exactly the modular weight 
to compensate this factor of $2M/N$.
The superpotential in Eq. (9) is not separable due to the different powers of
$\eta(T)$ in the two terms. This is a result of the modular weights of hidden 
matter and that of $detA$
which arises from the modular weights of  $\phi$.

Note that both the contribution of hidden matter condensates and the
dilaton independent term are essential to get a nonseparable superpotential. If there
is no hidden matter, target space dualtity requires that $h(T)=k(T)$ and
$W_{np}$ is separable. In this case the dilaton potential is stable but
$F_S=0$ at the minimum. On the other hand, if $c=0$ then the second term in
Eq. (9) vanishes and
$W_{np}$ is again separable. Only when both are present, there are two
terms with different moduli dependence as dictated by target space duality
and therefore the superpotential is not separable.

\bigskip
\centerline{\bf 3. The effective scalar potential}

One can absorb the moduli dependence of $detA$ into $h(T)$ and the
constant parts of $detA$ and $h(T)$ into $g(S)$ so that now
$$g(S)=-(Nk \phi^s)(32\pi^2e)^{M/N-1}exp(-32\pi^2S/N) \eqno(12)$$
and 
$$h(T)=\eta(T)^{2M/N-6-t/N} \eqno(13)$$
The superpotential in Eq. (9) is now given by
$$W_{np}=g(S)h(T)+ck(T) \eqno(14)$$
where $g(S)$ and $h(T)$ are given by Eqs. (12) and (13).
From the above formulas for the nonperturbative superpotential, the dilaton
F--term $F_S$ is
$$F_S={-1 \over {8S_R^{3/2}T_R^{3/2}}}\eta(T)^{d^\prime}
[(g(S)(1+2bS_R)+c\eta(T)^{d-d^{\prime}}] \eqno(15)$$
where $b=-32\pi^2/N$, $d=-6$ and $d^{\prime}=-6+2M/N+t/N$. 
The moduli F--term $F_T$ is
$$F_T={1 \over {4S_R^{1/2}T_R^{3/2}}}\eta(T)^{d^\prime}
[{G_2(T) \over 4\pi}(g(S)d^{\prime}+cd \eta(T)^{d-d^{\prime}})
-{3 \over {2T_R}}(g(S)+c \eta(T)^{d-d^{\prime}})] \eqno(16)$$
where $G_2(T)$ is the second Eisenstein function and arises due to
$\partial \eta(T)/ \partial T=-\eta(T) G_2(T)/4\pi$.
The effective scalar potential for $S$ and $T$ becomes
$$\eqalignno{V_{eff}&={1 \over 16S_RT_R^3}|\eta(T)^{d^\prime}|^2[|g(S)(1+2bS_R)
+c\eta(T)^{d-d^{\prime}}|^2 \cr
&+|{G_2(T) \over 4\pi}(g(S)d^{\prime}+cd \eta(T)^{d-d^{\prime}})
-{3 \over {2T_R}}(g(S)+c \eta(T)^{d-d^{\prime}})|^2 \cr
&-3|g(S)+c\eta(T)^{d-d^{\prime}}|^2] &(17)}$$
Due to the nonseparability of $W_{np}$ $V_{eff}$ is much more complicated
than in the separable superpotential case. In particular, the condition
for minimum in the dilaton direction $\partial V_{eff}/ \partial S=0$ 
is $S$ and $T$ dependent. This is in contrast to the case
with a separable superpotential in which the same condition 
depends only on $S$ and
not on $T$. This makes the analysis more difficult, e.g. one needs to
minimize $V_{eff}$ numerically as a function of four real variables
$S_R,S_I,T_R,T_I$ simultaneously.

For $W_{np}$ given by Eq. (14),  
$\partial V_{eff}/ \partial S$ is not
proportional to $F_S$ due to the nonseparable superpotential. Therefore,
generically one obtains $F_S\not=0$ in addition to $F_T\not=0$ at the
minimum of $V_{eff}$. 
This has been confirmed by numerical analysis
of $V_{eff}$. Thus, generically there is some amount of SUSY
breaking in the dilaton direction. However, this is not important if in
these cases $|F_T|>>|F_S|$ so that moduli breaking is the dominant effect.
The question we are interested in is whether
there are points in the parameter space of the model for which
$|F_S|>>|F_T|$.
In that case supersymmetry is mainly broken in the dilaton direction
in contrast to moduli SUSY breaking one obtains from nonseparable
superpotentials.

For the numerical analysis we need to choose some reasonable range
for the parameters of the model. For example, the constant $c$ 
which is crucial for dilaton supersymmetry
breaking also fixes the overall magnitudes of $F_S$ and $F_T$. Requiring
$TeV$ scale soft scalar masses or $m_{3/2}$ means that $c \sim
10^{-15}M_P^3$. We took $10^{-16}M_P^3<c<8 \times 10^{-15}M_P^3$ in our
numerical anlysis. $N$ which gives the hidden gauge group $SU(N)$
lies in the range, $2<N<8$, and the number of hidden matter multiplets
$M$ must satisfy $M<N$ for gaugino condensation to occur. $t$ which gives
the power of $\eta(T)$ in the determinant of the hidden mass matrix is
generically $O(10)$.  Hidden matter masses generically arise from nonrenormalizable 
higher order terms in the superpotential and so $s \sim O(5)$[\EA]. The parameter $t$
only appears in $d^{\prime}$ and therefore one can take $d^{\prime}$ to be the parameter
instead.The scalar VEVs which give masses to hidden matter
are fixed by the coefficient of the anomalous D--term[\DSW]
which is generic to string models to be $O(M_P/10)$[\CKM].

Even though our numerical search is not exhaustive, we find a large
number of points in the parameter space which gives a minimum of $V_{eff}$ with
$|F_S|>>|F_T|$. There
are enough points to be convinced  that this happens for a large 
and realistic range of model parameters. For example, for $N=3$, 
$d^{\prime}=-2$ and $c=5 \times 10^{-15}M_P^3$ the minimum of $V_{eff}$ is
at $S=0.29+i0.22$, $T=0.99+i1.00$ which gives the ratio
$R=|F_S|/|F_T|=53$. For $N=4$, $d^{\prime}=-5$ and $c=6 \times 10^{-15}M_P^3$
the minimum is at $S=0.30+i0.22$, $T=0.99+i1.00$ which gives
$R=713$ which is an order of magnitude larger than the previous case.
There are a large number of such points in the parameter space and these
two are given simply to demonstrate the case. Of course, there is also
a large part of the parameter space for which $|F_S|<<|F_T|$ and moduli
SUSY breaking is dominant.

The location of the minimum is quite robust at least in the range of
parameters we examined. For example, at the minimum of $V_{eff}$,
$S_R \sim 0.3$ which is somewhat smaller than the value required by
coupling constant unification. In additon, $T_R$ and $T_I$  at the minimum
turn out to be very close to $1$ in all cases with large $R=|F_S|/|F_T|$.
The dependence of the ratio $R$ on the different parameters of the model
such as $N,M,\phi,c,t$ etc. is very complicated.  In general it can be 
said that $|F_S|$ is inveresely proportional 
to $N$ whereas it is directly proportional to $c$. (We remind that for vanishing $c$ the 
superpotential is separable and $F_S=0$.) Note that
$N \geq 2$ and $c$ cannot be much larger than $10^{-15}M_P^3$ for
phenomenological reasons.
Finally, the cosmological constant is always of $O(m_{3/2}^2M_P^2)$ and
negative in the cases we examined for both dilaton and moduli 
SUSY breaking. One can make it vanish by simply fine tuning the 
constant $c$ which does not affect our results.

\bigskip
\centerline{\bf 4. Conclusions and discussion}

In this letter, we argued that one needs a nonseparable effective 
superpotential in order to get a nonzero dilaton F--term. 
If the superpotential is separable, then the minimum condition for the dilaton
automatically insures that $F_S=0$. We considered a simple but realistic
example of a nonseparable superpotential which was obtained by including
hidden matter condensation effects and a dilaton independent term in the
usual hidden gaugino condensation scenario. The superpotential is
nonseparable because it contains two terms with different modulus
dependence, one from condensation effects
and the other the dilaton independent term.
This is a result of the nonzero modular weights of hidden matter fields and
their masses. Due to the nonseparability
of the superpotential, $F_S\not=0$ in vacuum for generic values of the 
parameters of the model. The ratio $|F_S|/|F_T|$ depends on the specific
choice of the model parameters such as the hidden gauge group, the hiden matter
content, the hidden matter masses and the coefficient of the dilaton
independent term.

We gave two examples of such points in our results even though 
there are many more of them.
One can easily obtain $|F_S|$ which are one or
two orders of magnitude larger than $|F_T|$ and thus resulting in 
dominantly dilaton SUSY breaking. 
Even though our search of the parameter space was not
exhaustive, it clearly shows that once the effective superpotential is
not separable it is fairly easy to break SUSY in the dilaton
direction.

The superpotential was rendered nonseparable by including hidden matter
condensation effects and a dilaton independent term. Whereas the origin of
the former is evident the same is not true for the latter. It may arise
from the condensation of the antisymmetric rank two tensor present in all
string models or it may have some other origin. The crucial point for our purposes
is the different moduli dependence of the two terms in
$W_{np}$ which renders it nonseparable. As long as this is the case
our general conclusions remain valid.

The same ideas can also be explored in racetrack models for which there is
no need for a dilaton independent term since the dilaton is stabilized by
the interplay between the two gaugino condensation effects present. If the two
hidden gauge groups have different matter content with different masses
as expected in generic cases, the resulting effective superpotential will
not be separable. In this case, our results indicate that there will be
cases with dilaton SUSY breaking for some range of the parameters of the
model.


\vfill
\eject

\refout
\vfill
\eject

\end
\bye